\documentclass[subscriptcorrection,upint,varvw,barcolor=Goldenrod3,mathalfa=cal=euler,balance,hyphenate,french,pdf-a,nolists]{asmejour} %

\usepackage{xcolor}
\usepackage{tikz}
\usepackage{setspace}
\usetikzlibrary{arrows,shapes,positioning,shadows,trees,patterns,intersections,matrix,fit,backgrounds,chains}
\usepackage{multirow}
\usepackage{algorithm}
\usepackage[noend]{algpseudocode}
\usepackage{natbib}
\usepackage{ulem}
\usepackage{multirow}
\usepackage{booktabs}
\usepackage{subcaption}

\setcitestyle{citesep={,}}

\newcommand{\fig}{\text{Fig.~}}
\newcommand{\tab}{\text{Tab.~}}
\newcommand{\eq}{\text{Eq.~}}
\newcommand{\sez}{\text{Sec.~}}
\newcommand{\alg}{\text{Algorithm~}}
\newcommand{\app}{\text{Appendix~}}
\DeclareMathOperator*{\argmin}{arg\,min}
\DeclareMathOperator*{\maax}{max}

\hypersetup{%
	pdfauthor={John H. Lienhard},                       		   	
	pdftitle={ASME Journal Paper LaTeX Template},                  	
	pdfkeywords={ASME journal paper, LaTeX template, BibTeX style, asmejour class},
	pdfsubject = {Describes the asmejour LaTeX template},			
	pdflicenseurl={https://ctan.org/pkg/asmejour},
}

\JourName{Mechanical Design}

\begin{document}

\SetAuthorBlock{Gabriel Garayalde}{Department of Civil and Environmental Engineering, Politecnico di Milano,\\
Piazza L. Da Vinci 32, Milano, Italy \\
gabrielemilio.garayalde@polimi.it}

\SetAuthorBlock{Luca Rosafalco}{Department of Civil and Environmental Engineering, Politecnico di Milano,\\
Piazza L. Da Vinci 32, Milano, Italy \\
luca.rosafalco@polimi.it} 

\SetAuthorBlock{Matteo Torzoni\CorrespondingAuthor}{Department of Civil and Environmental Engineering, Politecnico di Milano,\\
Piazza L. Da Vinci 32, Milano, Italy \\
matteo.torzoni@polimi.it} 

\SetAuthorBlock{Alberto Corigliano}{Department of Civil and Environmental Engineering, Politecnico di Milano,\\
Piazza L. Da Vinci 32, Milano, Italy \\
alberto.corigliano@polimi.it} 

\title{Mastering truss structure optimization with tree search}

\keywords{Monte Carlo tree search, truss optimization, reinforcement learning, computational design synthesis}
   
\begin{abstract}
This study investigates the combined use of generative grammar rules and Monte Carlo Tree Search (MCTS) for optimizing truss structures. Our approach accommodates intermediate construction stages characteristic of progressive construction settings. We demonstrate the significant robustness and computational efficiency of our approach compared to alternative reinforcement learning frameworks from previous research activities, such as Q-learning or deep Q-learning. These advantages stem from the ability of MCTS to strategically navigate large state spaces, leveraging the Upper Confidence bounds for Trees formula to effectively balance exploitation-exploration trade-offs. We also emphasize the importance of early decision nodes in the search tree, reflecting design choices crucial for highly performative solutions. Additionally, we show how MCTS dynamically adapts to complex and extensive state spaces without significantly affecting solution quality. While the focus of this paper is on truss optimization, our findings suggest MCTS as a powerful tool for addressing other increasingly complex engineering applications.
\end{abstract}

\date{}

\maketitle 

\section{Introduction}

Machine learning (ML) is impacting engineering applications, from structural health monitoring \cite{CSAI22,SpringerChp22} and predictive maintenance \cite{Torzoni_DT}, to optimal flow control \cite{Fan2020} and automation in construction \cite{Elmaraghy2023}. Thanks to algorithmic advances and increased computational capabilities, there is a promise in enabling new approaches in computational design synthesis (CDS) --- a multidisciplinary research field aimed at automating the generation of design solutions for complex engineering problems \cite{AntonssonCagan2005,Chakrabarti2013,CampbellShea2014}. By integrating constraints related to the fabrication process, for instance through physics-based simulation, CDS could unlock the potential of additive manufacturing in various fields \cite{Wangler2016}, e.g., 3D concrete printing \cite{Giacomo2024}.

The effectiveness of traditional approaches to truss optimization, such as the ground-structure method \cite{Dorn1964,Rozvany1997}, has been established through decades of research. However, these methods suffer from high computational complexity and solution instability \cite{Gabriel,Ohsaki2011}. Alternative strategies for discrete truss optimization rely on heuristic techniques, including genetic algorithms \cite{Holland1992,Permyakov2006,HooshmandCampbell2016}, particle swarm \cite{KennedyEberhart1995,LuhLin2011}, differential evolution \cite{HoHuu2016}, and simulated annealing \cite{Lamberti2008}. Nevertheless, the applicability of these methods is similarly limited by their high computational burden and slow convergence as the size of the search space increases \cite{Ohsaki2011}. 

The search space of candidate solutions can be narrowed through generative design grammars \cite{Cagan2001}, which facilitate the exploration of alternative designs within a coherent framework \cite{MullinsRinderle1991,AlberRudolph2004}. These grammars are structured sets of rules that constrain the space of design configurations by accounting for mechanical information, such as the stability of static equilibrium. By integrating these rules within optimization procedures, it is therefore possible to explore incremental construction processes where the final design is reached through intermediate feasible configurations. The use of grammar-based approaches for truss topology generation and optimization has been proposed in \cite{ReddyCagan1995}, while their integration within heuristic approaches has been explored in \cite{Shea1997,SheaCagan1997,Puentes2020,Königseder2013,Fenton2015}. 

Recently, the optimal truss design problem has been formalized as a Markov decision process (MDP) \cite{Ororbia2021}. The solution to an MDP involves a series of choices, or actions, aimed at maximizing the long-term accumulation of rewards, which in this context measures the design objective. Viewing truss optimal design through the MDP lens, an action consists of adding or removing truss members, with the ultimate goal of optimizing a design objective, e.g., minimize the structural compliance. The final design thus emerges from a series of actions, possibly guided by grammar rules. This procedure is particularly suitable for truss structures, as it naturally accommodates discrete structural optimization, where adding a single member can significantly alter the functional objective of the design problem. Additionally, it can be extended to design optimization in additive manufacturing settings and continuum mechanics. The same methodology is similarly applicable to parametric optimization problems \cite{EJM2023}, including cases with stochastic control variables \cite{SciRep2023}.

Reinforcement learning (RL) is the branch of ML that addresses MDPs through repeated and iterative evaluations of how a single action affects a certain objective \cite{SuttonBarto2017}. Relevant instances of RL-based optimization in engineering include two-dimensional kinematic mechanisms \cite{Vermeer2018} and the ground-structure of binary trusses \cite{Zhu2021}. The advantage of RL over heuristic methods lies in its flexibility in handling high-dimensional problems, as demonstrated in \cite{Mazyavkina2021, Bello2016, JeonKim2020}. In \cite{Ororbia2021}, the MDP formalizing the optimal truss design has been solved using Q-learning \cite{Watkins1992}, constraining the search space with the grammar rules proposed in \cite{Lipson2008}. In a separate work \cite{Ororbia2023}, the same authors have also addressed the challenges of large and continuous design spaces through \textit{deep} Q-learning.

In this paper, we demonstrate how addressing optimal truss design problems with the Monte Carlo tree search (MCTS) algorithm \cite{Browne2012,LeventeSzepesvari2006} can offer significant computational savings compared to both Q-learning and deep Q-learning. MCTS is the RL algorithm behind the successes achieved by ``AlphaGo'' \cite{Silver2016} and its successors \cite{Silver2017,Schrittwieser2020} in playing board games and video games. In science and engineering, MCTS has been used for various applications employing its single-player games version \cite{Schadd2008}. Notable instances include protein folding \cite{Yang2017}, materials design \cite{Dieb2019,Dieb2017}, fluid-structure topology optimization \cite{Gaymann2019}, and the optimization of the dynamic characteristics of reinforced concrete structures \cite{Rossi2021}. 

For truss design, MCTS has been used in ``AlphaTruss'' \cite{Luo2022a} to achieve state-of-the-art performance while adhering to constraints on stress, displacement, and buckling levels. The same framework has been extended to handle continuous state-action spaces through either kernel regression \cite{Luo2022b} or soft actor-critic \cite{Du2023a} --- an off-policy RL algorithm. Despite the potential of using continuous descriptions of the design problem, the combination of RL and grammar rules proposed in \cite{Ororbia2021,Ororbia2023} remains highly competitive, as it enables constraining the design process with strong inductive biases reflecting engineering knowledge. Building on this insight, the novelty of our approach lies in the integration of MCTS with grammar rules to strategically navigate the solution space, allowing for significant computational gains compared to \cite{Ororbia2021,Ororbia2023}, where Q-learning and deep Q-learning have been respectively adopted. 

The effectiveness of the proposed approach lies in the MCTS capability to propagate information from the terminal nodes of the tree, which are associated with the final design performance, back to the ancestor nodes linked with the initial design states. This feedback mechanism allows for informing subsequent simulations, exploiting previously synthesized designs to enhance the decision-making process at initial branches and progressively refine the search towards optimal designs. Moreover, the probabilistic nature of MCTS enables the discovery of highly performative design solutions by balancing the exploitation-exploration trade-off. This balance is achieved through a heuristic hyperparameter that tunes the Upper Confidence bounds for Trees (UCT) formula, whose effect is investigated through a parametric analysis.

The remainder of the paper is organized as follows. Section \ref{sez:method} states the optimization problem and provides an overview of the MDP setting, the grammar rules, and the MCTS algorithm. In \sez\ref{sez:results}, the computational procedure is assessed on a series of case studies. We provide comparative results with respect to \cite{Ororbia2021,Ororbia2023}, demonstrating superior design capabilities, and test our methodology on two novel progressive construction setups. Section \ref{sez:conclusion} finally summarizes the obtained results and draws the conclusions.

\section{Methodology}
\label{sez:method}

In this section, we describe the methodology characterizing our optimal truss design strategy. This includes the physics-based numerical model behind the design problem in \sez\ref{sez:method_1}, the MDP formalizing the design process in \sez\ref{sez:method_2}, the grammar rules for truss design synthesis in \sez\ref{sez:method_3}, the MCTS algorithm for optimal truss design formulated as an MDP in \sez\ref{sez:method_4}, and the UCT formula behind the selection policy in \sez\ref{sez:method_5}, before detailing their algorithmic integration in \sez\ref{sez:method_6}.

\subsection{Optimal truss design problem}
\label{sez:method_1}
The design problem involves defining the truss geometry that optimizes a design objective under statically applied loading conditions. In the following, we consider minimizing the maximum absolute displacement experienced by the structure, although this is not a restrictive choice. This design setting, similar to the compliance minimization problem typical of topology optimization \cite{Gabriel}, has been retained for the purpose of comparison with \cite{Ororbia2021,Ororbia2023}.

For the sake of generality, we set the design problem in the context of continuum elasticity, of which truss design is an immediate specialization. Specifically, we seek a set of $I$ subdomains $\lbrace\Omega^s_1,\ldots,\Omega^s_I\rbrace$, each occupying a certain region of the design domain, whose union $\Omega=\bigcup^I_{i=1} \Omega^s_i$ minimizes the the structure’s displacement to the greatest extent, as follows:
\begin{equation}
     \bar{\Omega} =\bigcup^I_{i=1} \bar{\Omega}^s_i = \argmin_{\Omega=\bigcup^I_{i=1} \Omega^s_i} \lVert\mathbf{u}(\mathbf{x})\rVert_\infty, \quad \text{with } \mathbf{x}\in\Omega,
     \label{eq:minimization}
\end{equation}
where $\mathbf{u}$ is the displacement field, $\mathbf{x}$ are the spatial coordinates, and $\lVert\mathbf{u}\rVert_\infty$ is the infinity norm of $\mathbf{u}$, defined as $\lVert\mathbf{a}\rVert_\infty=\maax_{m}\lvert a_{m}\rvert$, with $a_{m}$, $m=1,\ldots,M$, being the $m$-th entry of $\mathbf{a}\in\mathbb{R}^{M}$. Problem \eqref{eq:minimization} is subjected to the following constraints, $\forall\Omega=\bigcup^I_{i=1} \Omega^s_i$:
    \begin{subequations}\label{eq:constrains}
    \begin{align}
        &\nabla \cdot \boldsymbol{\sigma} + \mathbf{b} = 0 &\text{equilibrium,} \\        
        &\boldsymbol{\sigma} = \mathbf{E} \boldsymbol{\epsilon} &\text{linear elastic constitutive law,} \\
        &\boldsymbol{\epsilon} = \frac{\nabla \mathbf{u} + \nabla \mathbf{u}^{\top}}{2} &\text{linear kinematic compatibility,}
    \end{align}
    \end{subequations}
ensuring the static elasticity condition. Herein, $\boldsymbol{\sigma}$ is the stress field; $\boldsymbol{\epsilon}$ is the strain field; $\mathbf{b}$ is the vector of body forces; $\mathbf{E}$ is the elasticity tensor; $(\nabla \cdot)$ is the divergence operator; and $(\nabla)$ is the gradient operator. 

Moreover, problem \eqref{eq:constrains} needs to be equipped with the following set of boundary conditions (BCs):
\begin{subequations}\label{eq:BC}
\begin{align}
        &\mathbf{u} = \mathbf{u}_g, &\text{on } \partial \Omega_g = \bigcup^I_{i=1} \partial {\Omega_g^s}_i \qquad~\text{Dirichlet BC,}\\
        &\boldsymbol{\sigma} \cdot \mathbf{n} = \mathbf{f}, &\text{on } \partial \Omega_h=\bigcup^I_{i=1} \partial {\Omega^s_h}_i \qquad\text{Neumann BC,}
\end{align}
\end{subequations}
where, $\partial \Omega_g$ and $\partial \Omega_h$ are the Dirichlet and Neumann boundaries, respectively; $\mathbf{u}_g$ is the assigned displacement field on $\partial \Omega_g$; $\mathbf{f}$ is the vector of surface tractions acting on $\partial \Omega_h$; and $\mathbf{n}$ is the outward unit vector normal to $\partial \Omega_h$. It is worth highlighting that this framework can be generalized to include non-linear constitutive behaviors --- an extension that will be explored in future works.

Equation \eqref{eq:minimization} can be easily adapted for planar trusses by introducing a finite element (FE) discretization to solve problem \eqref{eq:constrains}, defining each subdomain $\Omega^s_i$, $i=1,\ldots,I$, to be a truss element, with the union set operator representing the connections made through hinges. Accordingly, the optimization problem is reformulated as:
\begin{subequations}\label{eq:minimizationFE}
\begin{align}
\min_{\Omega=\bigcup^I_{i=1} \Omega^s_i} & \ \ \lVert\mathbf{U}(\Omega)\rVert_\infty, ~~ \text{with } \Omega^s_i \text{ a truss FE,} \\
\text{subject to}: & \ \ \mathbf{K}\mathbf{U} = \mathbf{F}, \quad ~ \text{in } \Omega=\bigcup^I_{i=1} \Omega_i,\label{eq:FE} \\
& \ \ \mathbf{U} = \mathbf{U}_0, ~ \quad \text{on } \partial\Omega_g = \bigcup^I_{i=1} \partial {\Omega_g^s}_i,\\
& \ \ V \leq V^{\text{max}}, \text{ with } V = \sum_{i=1}^I A_i L_i,
\end{align}
\end{subequations}
where $\mathbf{U}$ is the vector of nodal displacements; $\mathbf{K}$ is the stiffness matrix; $\mathbf{F}$ is the vector of forces induced by the external loadings; $\mathbf{U}_0$ is the vector of nodal displacements enforced on $\partial \Omega_g$; $A_i$ and $L_i$ are the cross-sectional area and the length of the $i$-th truss element $\Omega^s_i$, respectively; and $V^{\text{max}}$ is a prescribed threshold on the maximum allowed volume of the truss lattice. For further details on the FE method, the reader may refer to \cite{book:Belytschko00}.

\subsection{MDP framework for sequential decision problems}
\label{sez:method_2}

In a decision-making setting, an \textit{agent} must choose from a set of possible actions, each potentially leading to uncertain effects on the state of the system. The decision-making process aims to maximize, at least on average, the numerical utilities assigned to each possible action outcome. This involves considering both the probabilities of various outcomes and our preferences among them. 

In sequential decision problems, the agent's utility is influenced by a sequence of decisions. MDPs provide a framework for describing these problems in fully observable, stochastic \textit{environments} with Markov transition models and additive rewards \cite{Bellman1957}. Formally, an MDP is a $4$-tuple $\langle\mathcal{S},\mathcal{A},\mathcal{P},\mathcal{R}\rangle$, comprising a space of states $\mathcal{S}$ that the system can assume, a space of actions $\mathcal{A}$ that can be taken, a Markov transition model $\mathcal{P}$, and a space of rewards $\mathcal{R}$. The characterization of these quantities for truss optimization purposes is detailed below, after discussing their roles in MDPs.

We consider a time discretization of a planning horizon $(0,T)$ using non-dimensional time steps $t=0,\ldots,T$, and denote the system state at time $t$ as $s_t\in\mathcal{S}$, which is the realization of the random variable $S_t\sim p(s_t)$, with $p(s_t)$ being the probability distribution encoding the relative likelihood that $S_t = s_t$. Moreover, we denote the control input at time $t$ as $a_t\in\mathcal{A}$. The transition model \mbox{$\mathcal{P}:\mathcal{S}\times\mathcal{S}\times\mathcal{A}\mapsto[0,1]$} encodes the probability of reaching any state $s_{t+1}$ at time $t+1$, given the current state $s_t$ and an action $a_t$, i.e., $p(s_{t+1} | s_t, a_t)\in\mathcal{P}$. The reward $R_t\sim p(r_t)$, with $r_t\in\mathcal{R}$, quantifies the value associated with each possible set $\lbrace s_{t},a_{t},s_{t+1}\rbrace$.

We define a control policy $\pi:\mathcal{S}\mapsto\mathcal{A}$ as the mapping from any system state to the space of actions. The goal is to find the \textit{optimal} control policy $\pi^*(S_t)$ that provides the optimal action $a^*_t$ for each possible state $s_t$. The optimal policy $\pi^*(S_t)$ is learned by identifying the action $a^*_t$ that maximizes the expected utility over $(0,T)$. The problem of finding the optimal control policy is inherently stochastic. Consequently, the associated objective function is additive and relies on expectations \cite{papakonstantinou2014planning1}. This is typically expressed as the total expected discounted reward over $(0,T)$.

\begin{figure}[t]
\centering\includegraphics[width=0.62\linewidth]{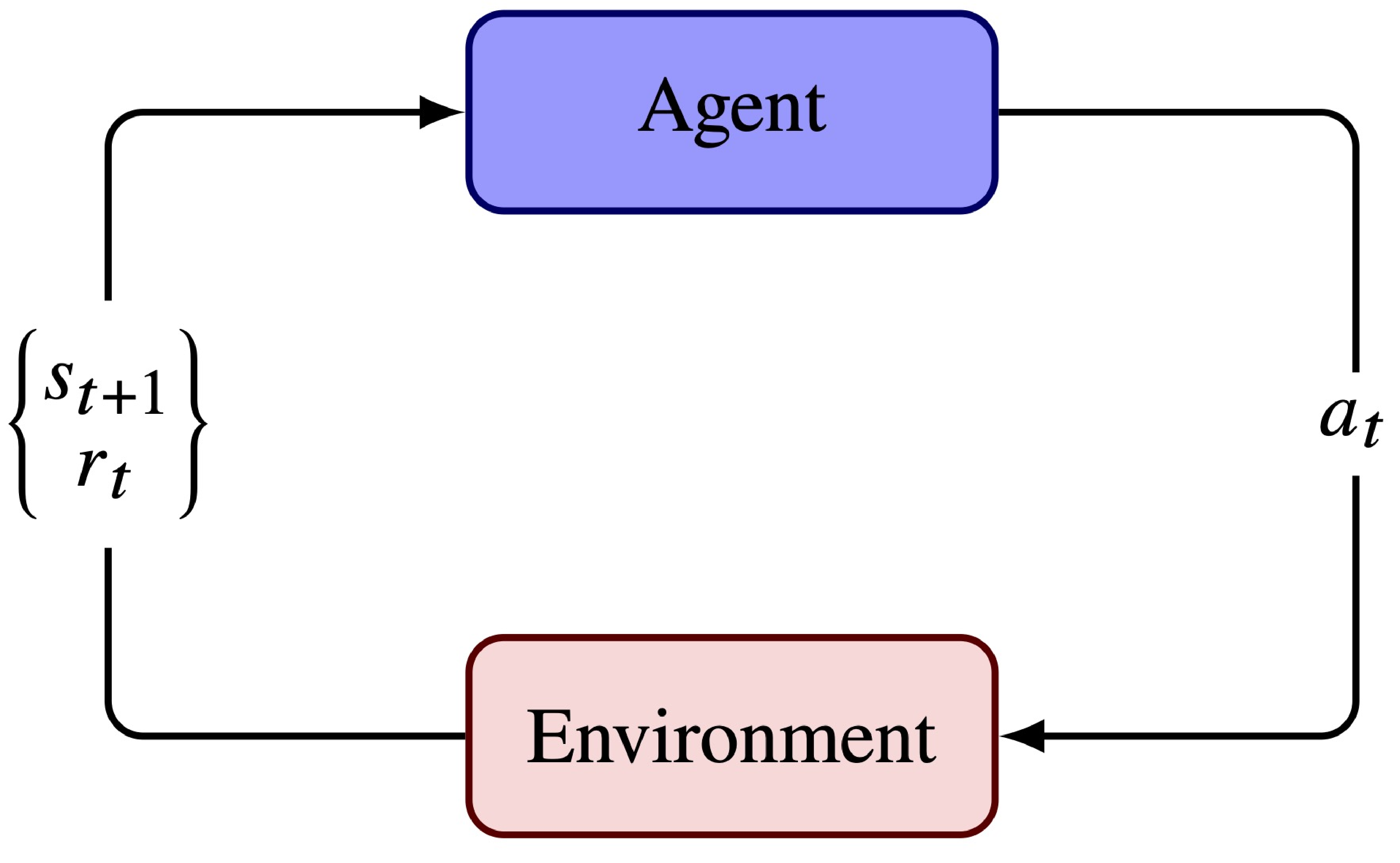}
\caption{Schematic agent-environment interaction.\label{fig:1}}
\end{figure}

The sequential decision problem can be viewed from the perspective of an agent-environment interaction, as depicted in \fig\ref{fig:1}. In this view, the agent perceives the environment and aims to maximize the long-term accumulation of rewards by choosing an action $a_t$ that influences the environment at time $t+1$. The environment interacts with the agent by defining the evolution of the system state, and providing a reward $r_t$ for taking $a_t$ and moving to $s_{t+1}$. 


One way to characterize an MDP is to consider the expected utility associated with a policy $\pi(S_t)$ when starting in any state $s_t$ and following $\pi(S_t)$ thereafter. To this aim, the \textit{state-value} function $V^\pi(S_t):\mathcal{S}\mapsto\mathbb{R}$ quantifies, for every state $s_t$, the total expected reward an agent can accumulate starting in $s_t$ and following policy $\pi(S_t)$. In contrast, the \textit{action-value} function, \mbox{$Q^\pi(S_t, a_t):\mathcal{S}\times\mathcal{A}\mapsto\mathbb{R}$} reflects the expected accumulated reward starting from $s_t$, taking action $a_t$, and then following policy $\pi(S_t)$. In both cases, the probability of each reaching any state $s_{t+1}$ is estimated using transition probabilities $p(s_{t+1}|s_t,a_t)$.

For our purposes of optimal truss design, we refer to a grid-world environment with a predefined number of nodes on which possible truss layouts can be defined. The reward function shaping $\mathcal{R}$ could account for local design objectives, such as the displacement at a prescribed node, or global performance indicators, such as the maximum absolute displacement, stress level, or strain energy. As previously commented, we monitor the maximum absolute displacement experienced by the structure. The state space $\mathcal{S}$ could potentially include any feasible truss layout resulting from progressive construction processes. Accordingly, the space of actions $\mathcal{A}$ could account for any possible modification of a given layout. In this scenario, the sizes of $\mathcal{S}$ and $\mathcal{A}$ increase significantly, even considering a reasonably small design domain. For this reason, explicitly modeling the Markov transition model $\mathcal{P}$ is not feasible.

The availability of a transition model for an MDP influences the selection of appropriate solution algorithms. Dynamic programming algorithms, for instance, require explicit transition probabilities. In situations where representing the transition model becomes challenging, a simulator is often employed to implicitly model the MDP dynamics. This is typical in episodic RL, where an environment simulator is queried with control actions to sample environment trajectories of the underlying transition model. Examples of such algorithms include Q-learning, as seen in \cite{Ororbia2021, Ororbia2023}, and MCTS, both of which approximate the action-value function and use this estimate as a proxy for the optimal control policy. As noted in \cite{Kocsis06}, convergence to the global optimal value function can only be guaranteed asymptotically in these cases. In our truss design problem, optimal planning is achieved via simulated experience provided by the FE model in \eq\eqref{eq:FE}, which can be queried to produce a sample transition given a state and an action.

\subsection{Grammar rules for truss design synthesis}
\label{sez:method_3}

To introduce the grammar rules that we employ to guide the process of optimal design synthesis, we refer to a starting seed configuration $s_0$, defined by deploying a few bars to create a statically determinate truss structure. This initial configuration must be modified through a series of actions selected by an agent. Every time an allowed action is enacted on the current state $s_t$, a new configuration $s_{t+1}$ is generated (see \fig\ref{fig:1}). The process continues until reaching a state $s_T$, characterized by a terminal condition, such as achieving the maximum allowed volume $V^{\text{max}}$ of the truss members.

\begin{figure}[t]
\centering\includegraphics[width=1\linewidth]{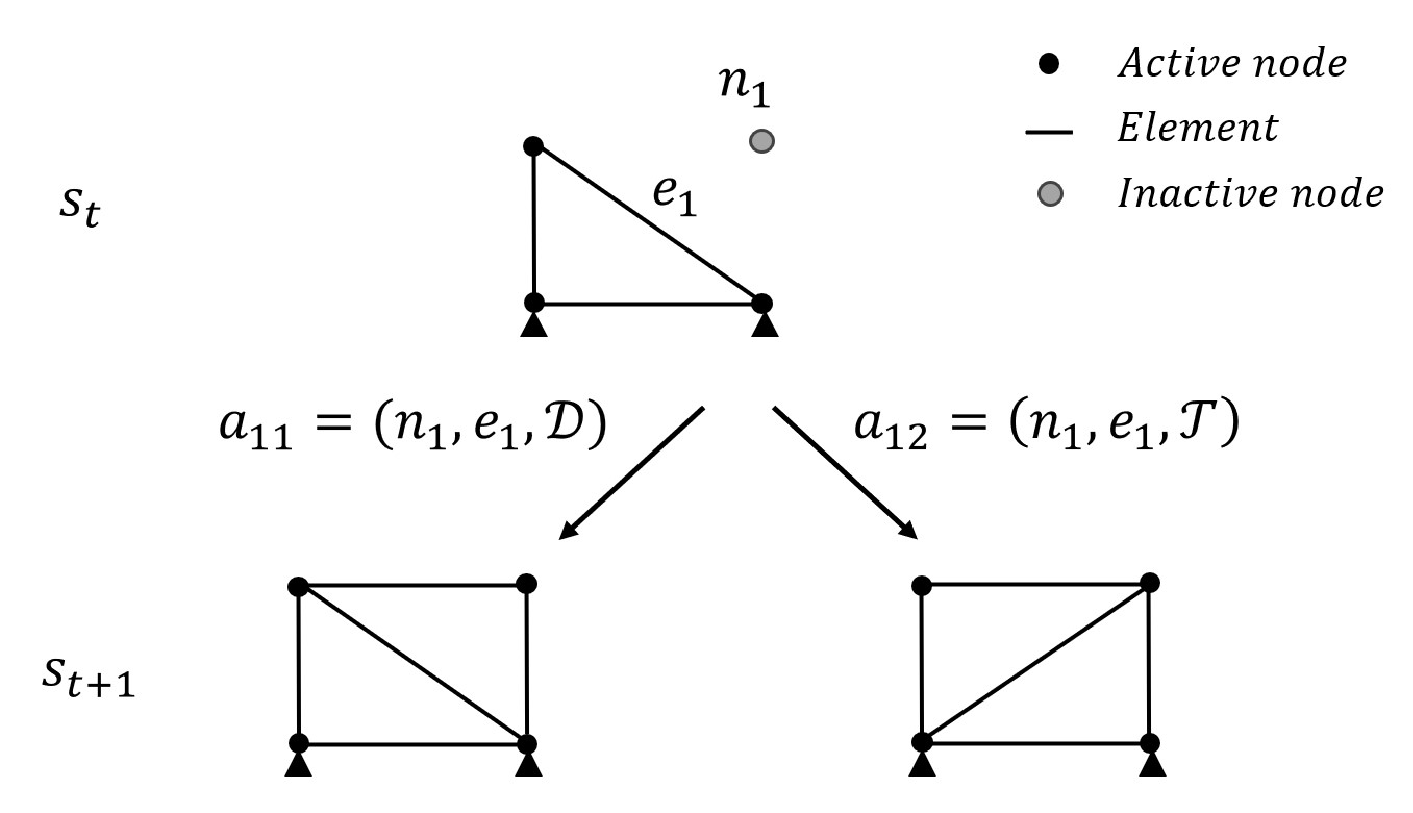}
\caption{Exemplary actions following operators $\mathcal{D}$ and $\mathcal{T}$. The current configuration $s_t$ (top) is modified either through action $a_1$ (bottom left) following the $\mathcal{D}$ operator or through action $a_2$ (bottom right) following the $\mathcal{T}$ operator, resulting in a new configuration $s_{t+1}$. In both cases, the selected truss element is $e_1$, and the chosen inactive node is $n_1$.\label{fig:2}}
\end{figure}

\begin{figure*}[ht]
\centering
\includegraphics[width=\textwidth]{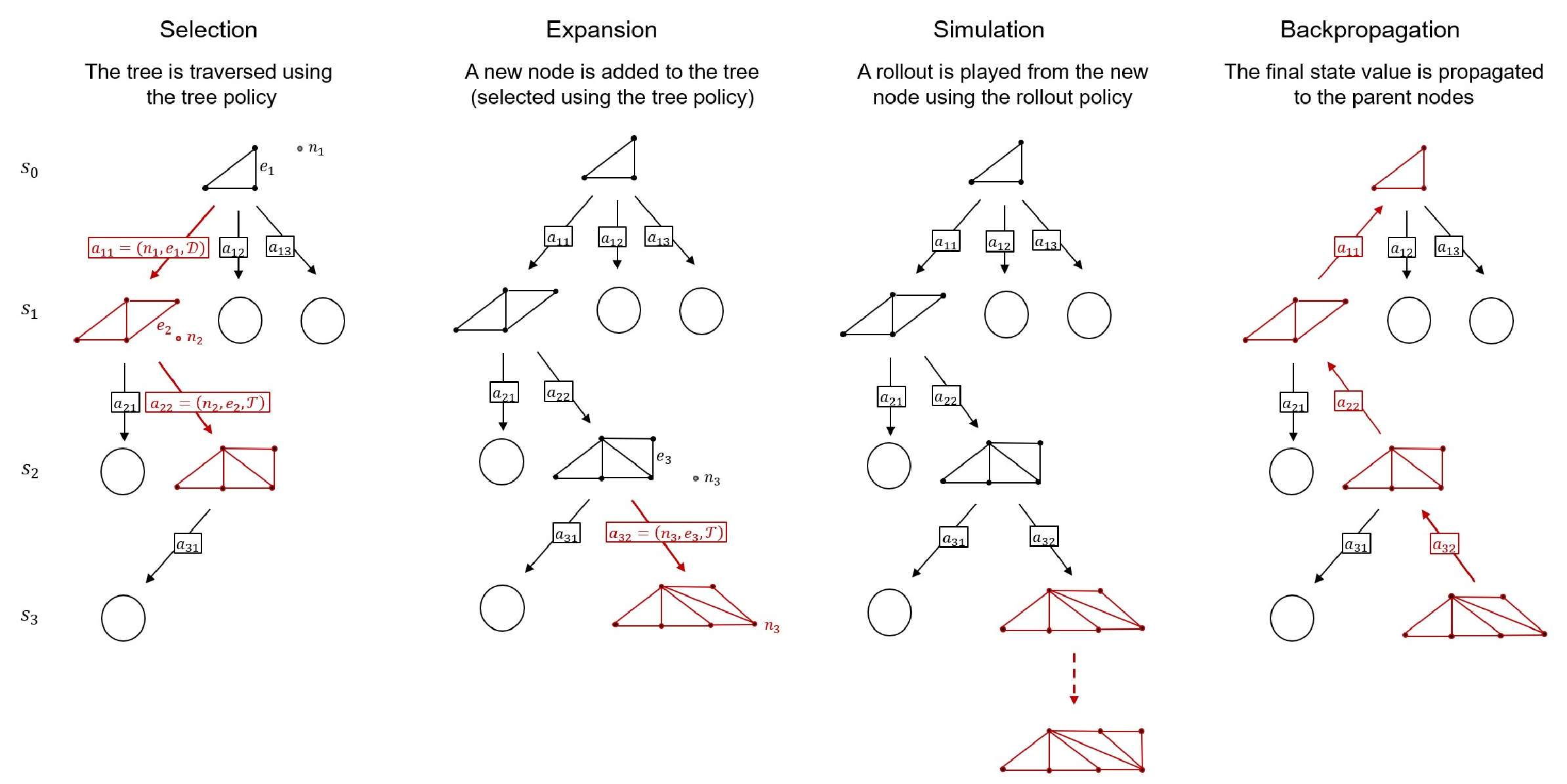}
\caption{Exemplary use of grammar rules for optimal truss design, formalized as a Markov decision process and solved through Monte Carlo tree search. The search tree construction and the corresponding truss design synthesis are achieved by repeating the four steps of selection, expansion, simulation, and backpropagation.\label{fig:3}}

\end{figure*}

To identify the allowed actions, we use the same grammar rules as those used in \cite{Lipson2008, Ororbia2021, Ororbia2023}. Starting from an isostatic seed configuration, these rules constrain the space of design configurations by allowing only truss elements resulting in triangular forms to be added to the current configuration, thereby ensuring statically determinate configurations. Given any current configuration $s_t$, an allowed action is characterized by a sequence of three operations:
\begin{enumerate}
\item Choosing a node among those not yet reached by the already placed truss elements. We term these nodes as \textit{inactive}, to distinguish them from the previously selected \textit{active} nodes.
\item Selecting a truss element already in place.
\item Applying a legal operator based on the position of the chosen node with respect to the selected element. The legal operators are either ``$\mathcal{D}$'' or ``$\mathcal{T}$'', see also \fig\ref{fig:2}. A $\mathcal{D}$ operator adds the new node and links it to the current configuration without removals, while a $\mathcal{T}$ operator also removes the selected element before connecting the new node. In both cases, the connections to the new node are generated ensuring no intersection with existing elements. 
\end{enumerate}

\subsection{Monte Carlo tree search}
\label{sez:method_4}

The MCTS algorithm is a decision-time planning RL method \cite{SuttonBarto2017}. It relies on two fundamental principles: ($i$) approximating action-values through random sampling of simulated environment trajectories, and ($ii$) using these estimates to inform the exploration of the search space, progressively refining the search toward highly-rewarding trajectories.

In the context of optimal truss design formulated as a sequential decision-making problem, MCTS incrementally grows a search tree where each node represents a specific design configuration, and edges correspond to potential state transitions triggered by allowed actions (see Figure~\ref{fig:3}). During training, the algorithm explores the search space of feasible truss designs to progressively learn a control policy, referred to as the \textit{tree policy}. This progressive policy improvement is based on value estimates of state-action pairs derived from previous runs of the algorithm, termed episodes. Each episode consists of four main phases \cite{SuttonBarto2017}:
\begin{enumerate}
    \item \textbf{Selection}: Starting from the root node associated with the seed configuration, the algorithm traverses the tree by selecting child nodes according to the tree policy until reaching a leaf node. The tree policy typically uses the UCT formula \cite{Kocsis06} to select child nodes. This formula ensures that actions leading to promising nodes are more likely to be chosen while still allowing for the exploration of less-visited nodes.  
    \item \textbf{Expansion}: If the selected leaf node corresponds to a non-terminal state $s_t$, the algorithm expands the tree by adding one or more child nodes representing unexplored actions from $s_t$. This expansion phase introduces new potential design configurations into the search tree, broadening the scope of exploration.
    \item \textbf{Simulation (Rollout)}: From one of the newly added nodes, the algorithm performs a path simulation or ``rollout'' to estimate the value gained by passing from that node. Since the tree policy does not yet cover the newly added nodes, MCTS employs a \textit{rollout policy} during this simulation phase to pick actions until reaching a terminal state $s_T$. The rollout policy is a random policy satisfying the truss design grammar rules, directing action along unexplored paths to backpropagate the associated reward signal back up the decision tree. While the tree policy expands the tree via selection and expansion, the rollout policy simulates environment-interaction based on random exploration. 
    \item \textbf{Backpropagation}: Upon reaching a terminal state $s_T$, the associated design is synthesized to evaluate the design objective. This reward is then backpropagated through the nodes traversed during selection and expansion. This process involves updating the visit counts of the nodes and the values $Q^\pi(s,a)$ for the corresponding state-action pairs, both of which influence the decision-making process through the UCT formula, as detailed in the following section. 
\end{enumerate}
Each time a reward signal is backpropagated to update the action-value estimates of state-action pairs, an episode is completed. This iterative process progressively refines the tree policy, making actions that lead to better rewards more likely to be chosen in future episodes, while still allowing for the exploration of new design configurations. The number of episodes is determined by the available computational budget. As the number of episodes increases, more nodes are added to the tree, and the precision of the Monte Carlo estimates for the mean return from each state-action pair improves. After completing the prescribed number of episodes, a deterministic policy can be derived by selecting, for example, the action with the highest estimated value $Q^\pi(s,a)$ at each state.

The advantages of MCTS stem from its online, incremental, sample-based value estimation and policy improvement. MCTS is particularly adept at managing environments where rewards are not immediate, as it effectively explores broad search spaces despite the minimal feedback. This makes MCTS especially suitable for progressive construction settings, where the final design requirements often differ from those of intermediate structural states. Intermediate construction stages typically involve sustaining self-load only, while different combinations of dead and live loads are experienced during operations. This capability stems from the backpropagation step, which allows information related to $s_T$ to be transferred to the early nodes of the tree. In contrast, bootstrapping methods like Q-learning may require a longer training phase to equivalently backpropagate information, as we demonstrate in \sez\ref{sez:results}. Further advantages of MCTS include: ($i$) accumulating experience by sampling environment trajectories, without requiring domain-specific knowledge to be effective; ($ii$) incrementally growing a lookup table to store a partial action-value function for the state-action pairs yielding highly-rewarding trajectories, without needing to approximate a global action-value function; ($iii$) updating the search tree in real-time whenever the outcome of a simulation becomes available, in contrast, e.g., with minimax's iterative deepening; and ($iv$) focusing on promising paths thanks to the selective process, leading to an asymmetric tree that prioritizes more valuable decisions. This last aspect not only enhances the algorithm's efficiency but can also offer insights into the domain itself by analyzing the tree's structure for patterns of successful courses of action.

\subsection{Upper Confidence bounds for Trees}
\label{sez:method_5}

The UCT formula is widely used as a selection policy in MCTS due to its ability to balance exploitation and exploration. In this work, we employ a modified UCT formula, compared to the one proposed in \cite{LeventeSzepesvari2006}, by introducing an $\alpha$ parameter that scales the relative weights of the exploitation and exploration terms as follows:
\begin{equation}
\label{eq:UCT}
\text{UCT}_j = (1-\alpha)\frac{v^\Sigma_j}{n_j} + \alpha \sqrt{\frac{2\log \sum_l n_l}{n_j}},
\end{equation}
which provides the UCT score of the $j$-th child node of $s_t$. Herein: $v^\Sigma_j$ is the Monte Carlo estimate of the total return gained by passing through the $j$-th child node, where this return represents the sum of all the terminal state rewards $r_T$ achieved after traversing the $j$-th child node; $n_j$ is the number of episode runs passing through the $j$-th child node; $\sum_l n_l$ is the total number of episode runs traversing the children of $s_t$; and $\alpha$ is a parameter that balances exploitation (average reward for the $j$-th child node) and exploration (encouraging exploration of nodes that have been visited less frequently than their siblings), respectively encoded in the first and second terms. It is also worth noting that both $v^\Sigma_j$ and $n_j$ are updated after each training episode.

\subsection{Algorithmic description}
\label{sez:method_6}

The algorithmic description of the optimal truss design strategy using the proposed MCTS approach is detailed in \alg\ref{al:algorithm}. It begins by initializing the root node with a seed configuration and then iteratively explores potential truss configurations through a sequence of selection, expansion, simulation, and backpropagation phases. In each episode, the algorithm selects a child node based on the UCT formula, generates and evaluates a new child node from a possible action, simulates random descendant nodes to explore the design space, and backpropagates the computed reward to update the policy.

\begin{algorithm}[ht]
\hspace*{\algorithmicindent} \textbf{input}: 
number of episodes $N_e$\\
\hspace*{40.2pt} parametrization of the physics-based model\\
\hspace*{40.2pt} grid design domain\\
\hspace*{40.2pt} seed configuration\\
\hspace*{40.2pt} grammar rules for truss design synthesis\\
\hspace*{40.2pt} exploration parameter $\alpha$
\begin{algorithmic}[1]
\State initialize root node for the seed configuration
\For {$N_e$}
\State $t = 0$
\State set root node for $s_{t=0}$ (seed configuration)
\Statex\hfill\Comment{selection}
\While {$t < T$ and $s_t$ previously explored}
    \State select $s_{t+1}$ via UCT formula
    \State $t\leftarrow t+1$
\EndWhile
\Statex\hfill\Comment{expansion}
\If {$t < T$ and $s_t$ not previously explored} 
    \For {states $s_{t+1}$ from allowed actions $a_t$}
        \State solve static equilibrium for $s_{t+1}$
        \State compute design objective $\lVert\mathbf{U}(\Omega)\rVert_\infty$
    \EndFor
    \State $t\leftarrow t+1$

\EndIf
\Statex\hfill\Comment{simulation}
\While {$t < T$}
    \State select a random child $s_{t+1}$
    \If {$s_{t+1}$ not previously explored}
        \State solve static equilibrium for $s_{t+1}$
        \State compute design objective $\lVert\mathbf{U}(\Omega)\rVert_\infty$
    \EndIf
    \State $t\leftarrow t+1$
\EndWhile
\Statex\hfill\Comment{backpropogation}
\State compute reward $r_T$ from terminal state $s_T$
\While {$t>0$}
    \State append $r_T$ to $s_{t}$ rewards list
    \State $s_{t}$ visit count += 1
    \State $t\leftarrow t-1$
\EndWhile
\EndFor
\State \Return deterministic control policy $\pi \approx \pi^*$
\end{algorithmic}
\caption{Monte Carlo tree search for optimal truss design.\label{al:algorithm}}
\end{algorithm}

\section{Results}
\label{sez:results}
In this section, we assess the proposed MCTS framework on different truss optimization problems. First, we adopt six case studies from \cite{Ororbia2021, Ororbia2023}, each featuring different domain and boundary conditions, to directly compare the achieved performance. Then, we consider two additional case studies to demonstrate the applicability of our procedure for progressive construction purposes. While in the former case studies the seed configuration fully covers the available design domain, in the latter we allow the seed configuration to grow --- mimicking an additive construction process ---  until reaching a terminal node at the far end of the domain.

The experiments have have been implemented in \texttt{Python} using the \texttt{Spyder} development environment. All computations have been carried out on a PC featuring an \texttt{AMD Ryzen\textsuperscript{TM} 9 5950X} CPU @ 3.4 GHz and 128 GB RAM.

\subsection{Truss optimization}

In the following, we present the results achieved for the six case studies adapted from \cite{Ororbia2021,Ororbia2023}, providing comparative insights for each scenario. All case studies deal with planar trusses, with truss elements featuring dimensionless Young's modulus $E=10^3$ and cross-sectional area $A=1$. The applied forces have a dimensionless value of $f_x = f_y = 10$, as per \cite{Ororbia2021,Ororbia2023}. The \textit{monitored} displacement refers to the maximum absolute displacement experienced by the structure.

\begin{figure*}[ht]
\centering
\includegraphics[width=1\textwidth]{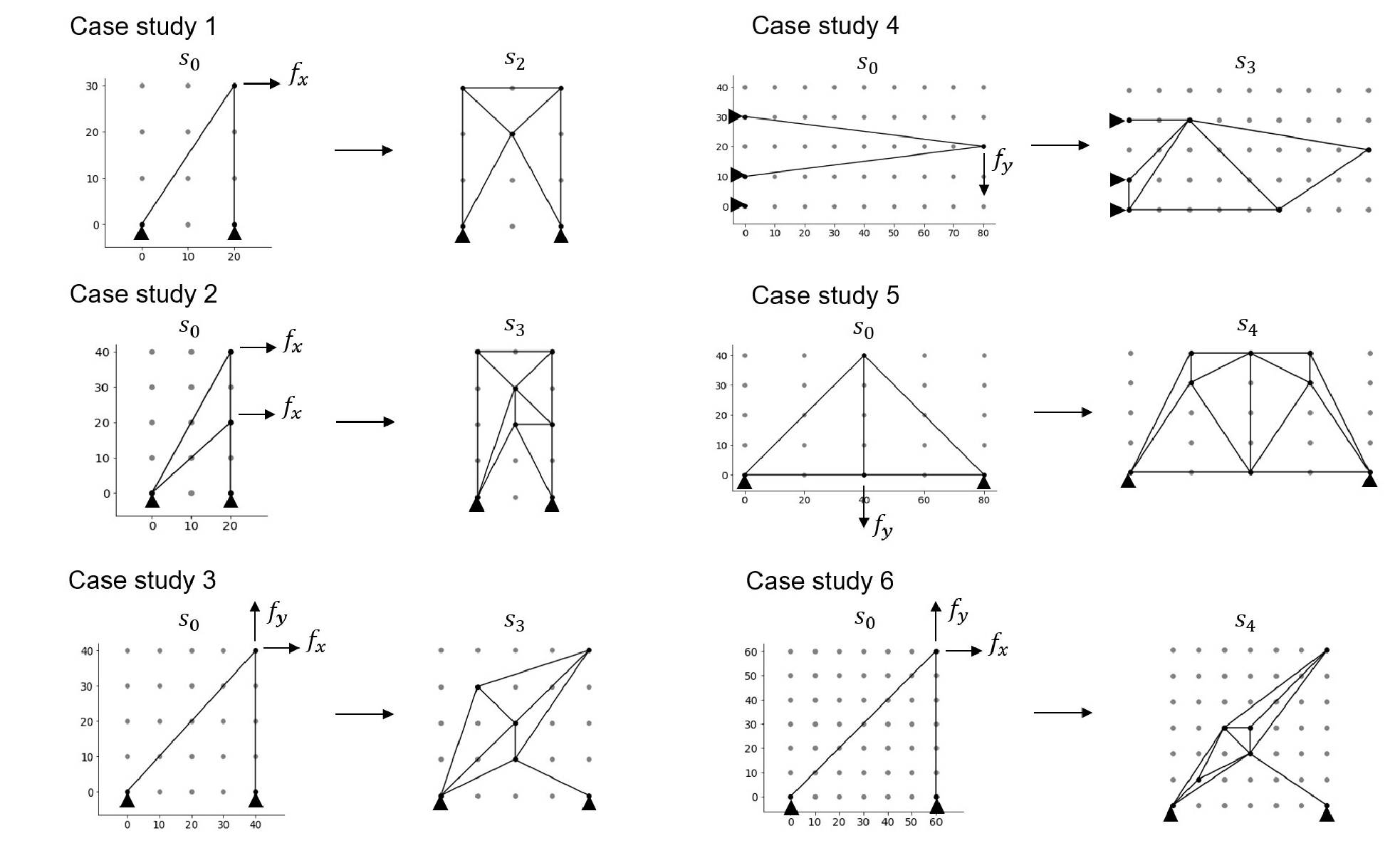}
\caption{Truss optimization - Case studies adapted from \cite{Ororbia2023}: summary of design domain, seed configuration $s_0$, and target optimal design $s_T$ identified through a brute-force exhaustive search.\label{fig:examples1-6}}
\end{figure*}

Each row of \tab\ref{tab:case_study_results} describes a case study in terms of design domain size, number of decision times or planning horizon $T$, and volume threshold $V^{\text{max}}$. These parameters have been set according to \cite{Ororbia2021,Ororbia2023} to facilitate the comparison between the proposed MCTS procedure and the Q-learning methods. For each case study, the design domain, structural seed configuration, externally applied force(s), and boundary conditions are shown under the corresponding $s_0$ label in \fig\ref{fig:examples1-6}. The target \textit{optimal} configuration $s_T$, identified through a brute-force exhaustive search of the state space, is illustrated under the $s_T$ label. Case Study 4 is the only one that differs from the reference due to the additional constraint at $(0,0)$.

\begin{table}[t]
\centering
\caption{Truss optimization - Problems settings description.\label{tab:case_study_results}}
\begin{tabular}{c c c c}
\toprule
 & \textbf{Domain size} & \textbf{Decision times} $T$ & \textbf{$V^{\text{max}}$ threshold}\\
\hline
Case 1 & $4\times3$ & $2$ & $160$ \\
Case 2 & $5\times3$ & $3$ & $240$ \\
Case 3 & $5\times5$ & $3$ & $225$ \\
Case 4 & $5\times9$ & $3$ & $305$  \\
Case 5 & $5\times5$ & $4$ & $480$  \\
Case 6 & $7\times7$ & $4$ & $350$  \\
\bottomrule
\end{tabular}
\end{table}

\begin{figure*}[ht]
\centering
\includegraphics[width=1\textwidth]{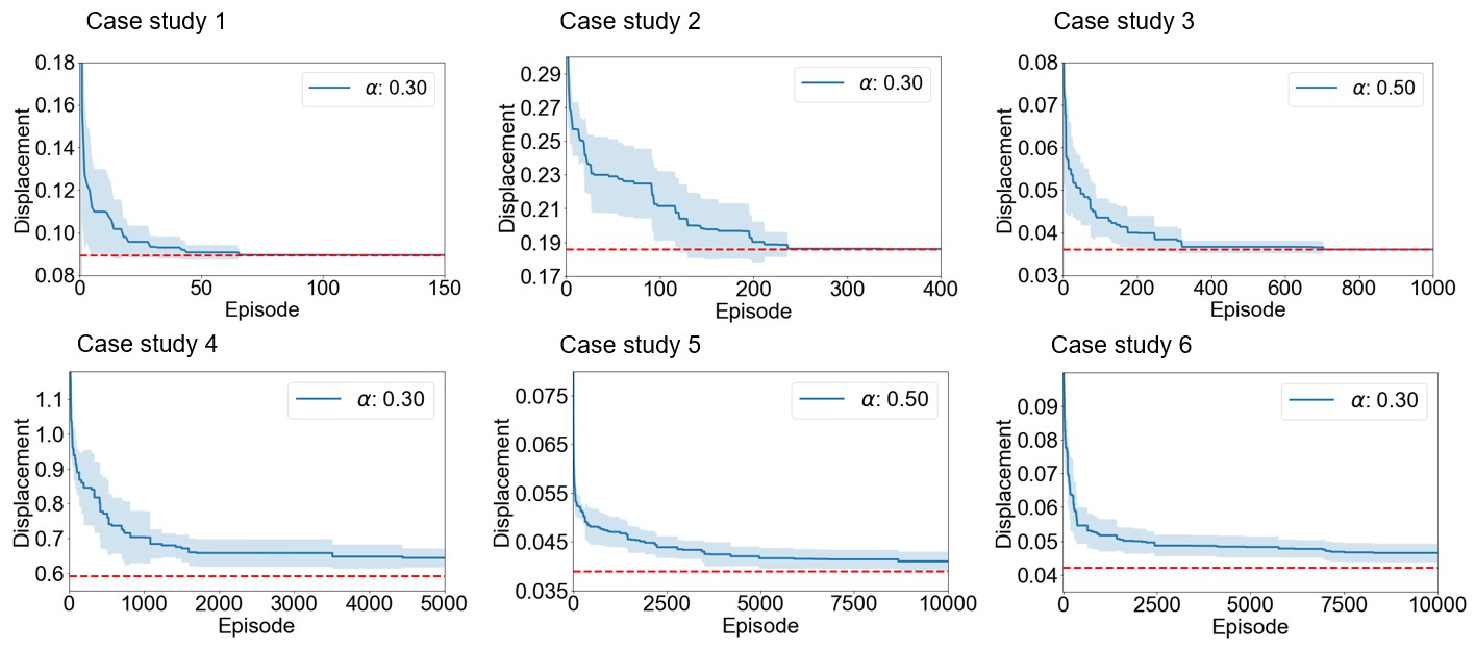}
\caption{Truss optimization - Case studies 1-6: evolution of the design objective during training, shown as the average value (solid blue line) with its one-standard-deviation credibility interval (shaded blue area), and target global minimum (dashed red line). Results averaged over 10 training runs.\label{fig:examples1-6_displacement}}
\end{figure*}

For each case study, \fig\ref{fig:examples1-6_displacement} shows the evolution of the design objective, i.e., the maximum absolute displacement experienced by the structure, as the number of training episodes increases. Results are reported in terms of average displacement (solid blue line) and one-standard-deviation credibility interval (dashed blue area), over 10 independent training runs. Each run utilizes MCTS for a predefined number of episodes. In practice, the number of episodes is set after an initial long training run in which we assess the number of episodes required to achieve convergence --- which typically depends on the complexity of the case study. After each training run, the best configuration is saved to subsequently compute relevant statistics. The attained displacement values are compared with those associated with the global minima (red dashed lines), representing the optimal design configurations in \fig\ref{fig:examples1-6}.

\begin{table}[t]
\caption{Truss optimization - Case studies 1-6: optimal design objective $\lVert\mathbf{U}(\overline{\Omega})\rVert_\infty$, percentage ratio of the optimal design objective to the displacement achieved by the learned policy, percentile score relative to the exhaustive search space, number of finite element evaluations required to achieve a near-optimal or optimal policy, and relative speed-up compared to \cite{Ororbia2023}. The speed-up is not reported for case study 4, as it differs from the reference for the additional constraint at $(0,0)$. Results averaged over 10 training runs.
 \label{tab:case_study_percentile_scores}}
\small
\centering{%
\setlength{\tabcolsep}{3pt} 
\begin{tabular}{ccccccc}
\toprule
& \multirow{2}{*}{$\lVert\mathbf{U}(\overline{\Omega})\rVert_\infty$} & \textbf{Objective} &\textbf{Percentile} & \multirow{2}{*}{\textbf{FE runs}} & \multirow{2}{*}{\textbf{FE runs vs \cite{Ororbia2023}}}\\
& &\textbf{ratio} & \textbf{score} & &\\
\hline
Case 1 & $0.0895$ & $100\%$   & $100\%$   & $106$ & $-74.70\%$\\
Case 2 & $0.1895$ & $100\%$   & $100\%$   & $517$ & $-76.27\%$\\
Case 3 & $0.0361$ & $100\%$   & $100\%$   & $966$ & $-56.51\%$\\ 
Case 4 & $0.5916$ & $91.91\%$ & $99.90\%$ & $1672$ & $\text{N/A}$\\
Case 5 & $0.0390$ & $95.23\%$ & $99.99\%$ & $9739$ & $-70.74\%$\\
Case 6 & $0.0420$ & $90.44\%$ & $99.98\%$ & $7931$ & $-31.25\%$\\
\bottomrule
\end{tabular}
}%
\end{table}

\begin{figure}[t]
\centering
\includegraphics[width=\linewidth]{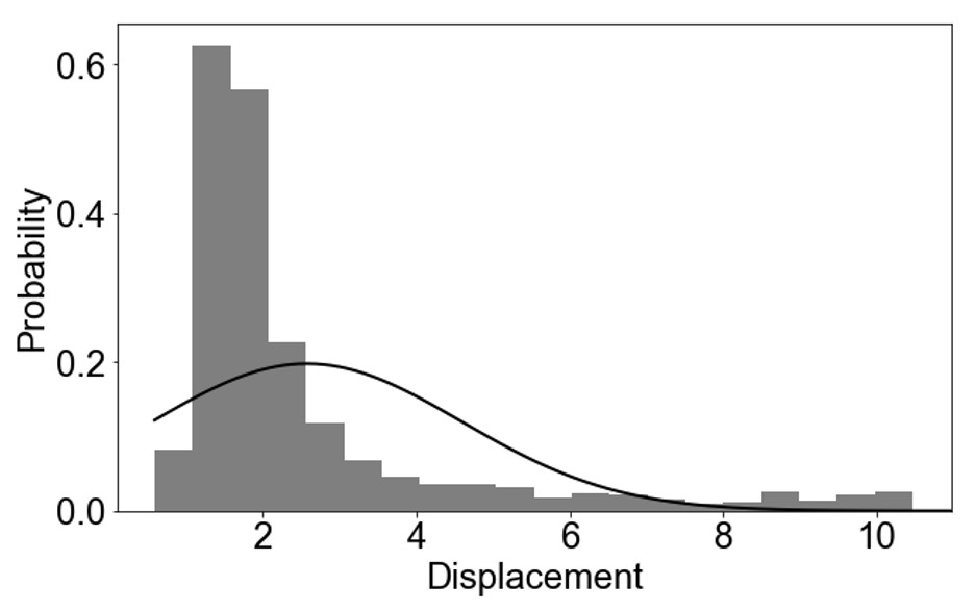}
\caption{Truss optimization - Case study 4: design objective distribution over the population of designs synthesized from an exhaustive search of the state space.\label{fig:example4_exhaustive}}
\end{figure}

The heuristic $\alpha$ parameter in \eq\eqref{eq:UCT} controls the balance between exploitation and exploration. The $\alpha$ values employed for the six case studies are overlaid on each learning curve in \fig\ref{fig:examples1-6_displacement}. Since an optimal value for this parameter is not known a-priori, this is set using a rule of thumb derived through a parametric analysis, as explained in the following section for case study 4.

A quantitative assessment of the optimization performance for each case study is summarized in \tab\ref{tab:case_study_percentile_scores}. Results are reported in terms of the optimal design objective $\lVert\mathbf{U}(\overline{\Omega})\rVert_\infty$, the percentage ratio of the optimal design objective to the displacement achieved by the learned policy, and the percentile score relative to the exhaustive search space. To clarify, a percentile score of $100\%$ corresponds to reaching the global optimum. A lower score, such as $99\%$, indicates that the design objective achieved with the final design $s_T$, synthesized from the learned optimal policy, is lower than the displacement associated with $99\%$ of all the possible configurations explored through an exhaustive search. An exemplary distribution of the design objective across the population of designs synthesized from the exhaustive search of the state space is shown in \fig\ref{fig:example4_exhaustive} for case study 4. Interestingly, the distributions obtained for the other case studies also exhibit a lognormal-like shape, although these are not shown here due to space constraints. While the objective ratio provides a dimensionless measure of how close the achieved design is to the global optimum in terms of performance, the percentile score quantifies the capability of MCTS to navigate the search space and find a design solution close to the optimal one. Both performance indicators are computed by averaging over 10 training runs. Additionally, we report the number of FE evaluations required to achieve a near-optimal or optimal policy, also averaged over 10 training runs, and indicate the percentage savings in the number of FE evaluations compared to those required by the deep Q-learning strategy from \cite{Ororbia2023}. It is worth noting that FE evaluations are only performed for the terminal state $s_T$, after it has been selected.

\begin{figure*}[!ht]
\centering
\includegraphics[width=1\textwidth]{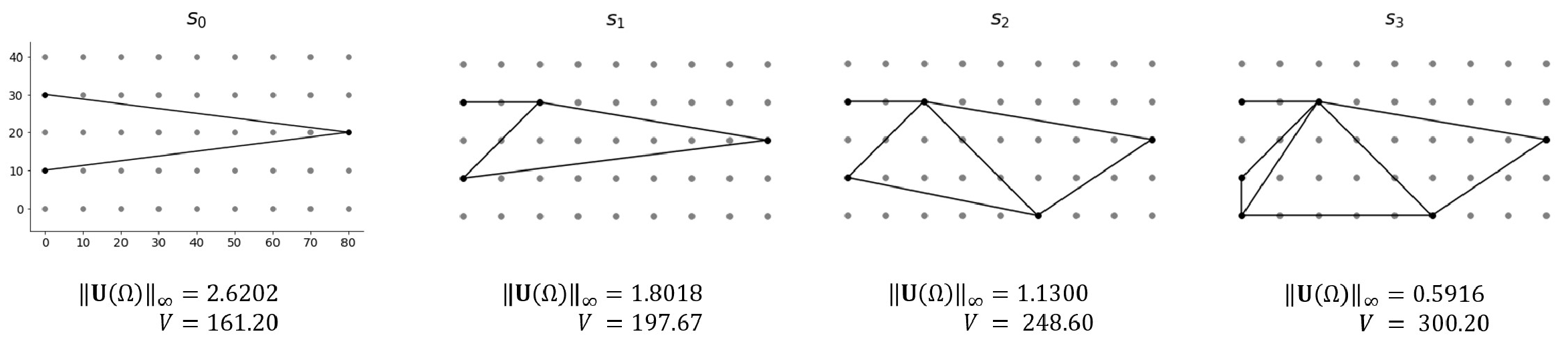}
\caption{Truss optimization - Case study 4: sequence of design configurations from the target optimal policy, identified through a brute-force exhaustive search, with details about the design objective value and the truss lattice volume.\label{fig:example4_states}}
\end{figure*}

\begin{figure*}[!ht]
\begin{subfigure}{.5\textwidth}
  \centering
  \includegraphics[width=.95\textwidth]{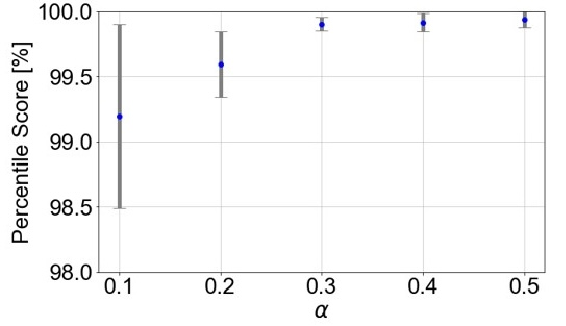}
  \caption{\label{fig:fig8a}}
\end{subfigure}%
\begin{subfigure}{.5\textwidth}
  \centering
  \includegraphics[width=.95\textwidth]{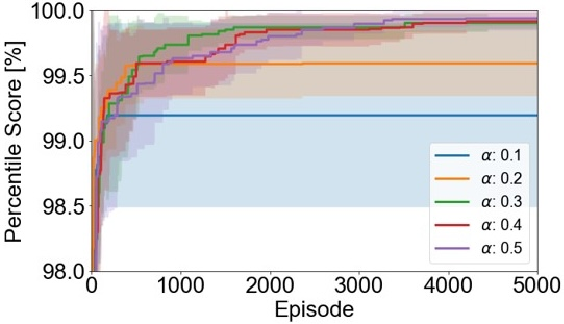}
  \caption{\label{fig:fig8b}}
\end{subfigure}%
\caption{Truss optimization - Case study 4: impact of varying the $\alpha$ parameter on the attained percentile score relative to the exhaustive search space. For each value of $\alpha$, results are reported in terms of (a) the average percentile score with its one-standard-deviation credibility interval, and (b) the evolution of the percentile score during training, shown as the average value with its credibility interval. Results averaged over 10 training runs.\label{fig:example4_percentile}}
\end{figure*}

\begin{figure}[ht]
\centering
\includegraphics[width=\linewidth]{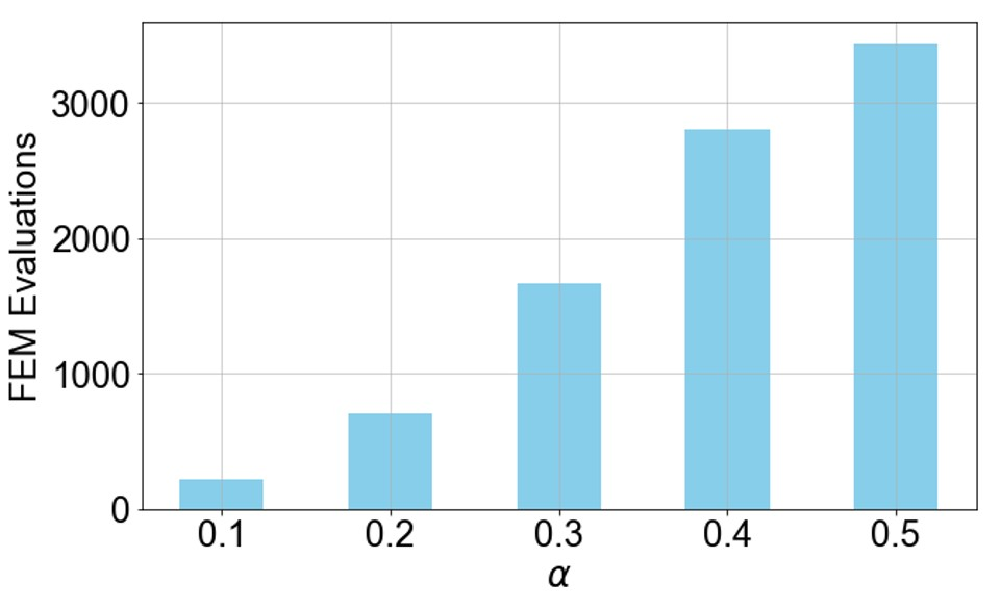}
\caption{Truss optimization - Case study 4: number of finite element evaluations required to achieve a near-optimal design policy for varying values of the $\alpha$ parameter.\label{fig:example4_FEM_alpha}}
\end{figure}

\subsection{Case study 4 --- detailed analysis}
\label{sez:example_4}

In this section, we provide a detailed analysis of case study 4. We select this case study because it is the only one in which we employ boundary conditions different from those reported in \cite{Ororbia2023}, which is useful for checking the MCTS capability to exploit constrained domain portions not in the seed configuration. Figure \ref{fig:example4_states} illustrates the sequence of structural configurations synthesized from the optimal policy obtained through exhaustive search. For each decision time, we report the corresponding value of the design objective and the volume of the truss lattice below the synthesized configuration $s_t$.

Figure \ref{fig:example4_percentile} summarizes the impact of varying the $\alpha$ parameter on the attained percentile score, to provide insights into the selection of an appropriate value. Specifically, \fig\ref{fig:fig8a} shows the percentile score relative to the exhaustive search space for the different $\alpha$ values, averaged over 10 training runs. Figure \ref{fig:fig8b} illustrates how the percentile score evolves as the number of episodes increases, offering insights into the effect of $\alpha$ on the convergence of MCTS. To compare the achieved performance for varying $\alpha$ values with the associated computational burden, \fig\ref{fig:example4_FEM_alpha} presents the number of FE evaluations required to achieve a near-optimal design policy, revealing an almost linear increase in the number of FE evaluations as $\alpha$ grows. Therefore, we consider $\alpha = 0.3$ to provide an appropriate balance between exploitation and exploration, yielding an average percentile score of $99.90\%$  across 10 training runs, which is close to the scores for $\alpha=0.4$ and $\alpha=0.5$, but with only 1672 FE evaluations. The achieved ratio of the optimal design objective to the displacement achieved by the learned policy is $91.91\%$ (see \tab\ref{tab:case_study_percentile_scores}). Similar results from the parametric analysis of $\alpha$ for the other case studies are provided in \app\ref{sec:parametric}.

\subsection{Progressive construction}
\label{sez:cantilever}

In this section, we showcase the potential of the proposed MCTS strategy in guiding the progressive construction of a truss cantilever beam and a bridge-like structure. Unlike in the previous case studies, where a simplified seed configuration was initially assigned to comply with the target boundary conditions and then refined, here we allow the seed configuration to progressively grow until reaching a prescribed terminal node not included in the initial configuration. Therefore, the agent must account for the intermediate construction stages per se, not just as necessary steps to reach the final configuration. Another difference compared to the previous case studies is that instead of considering a fixed loading configuration, the structure is subjected to self-weight (unit dimensionless density), modifying the loading configuration at each stage. However, as in the previous cases, since the design process aims to maximize the performance of the final configuration, the chosen design objective is again the maximum absolute displacement. Although we did not set a limit on the maximum number of states, the agent must strike a balance between achieving higher structural stiffness by adding additional members and the weight these extra elements bring. 

\begin{figure*}[t]
\centering
\includegraphics[width=1\textwidth]{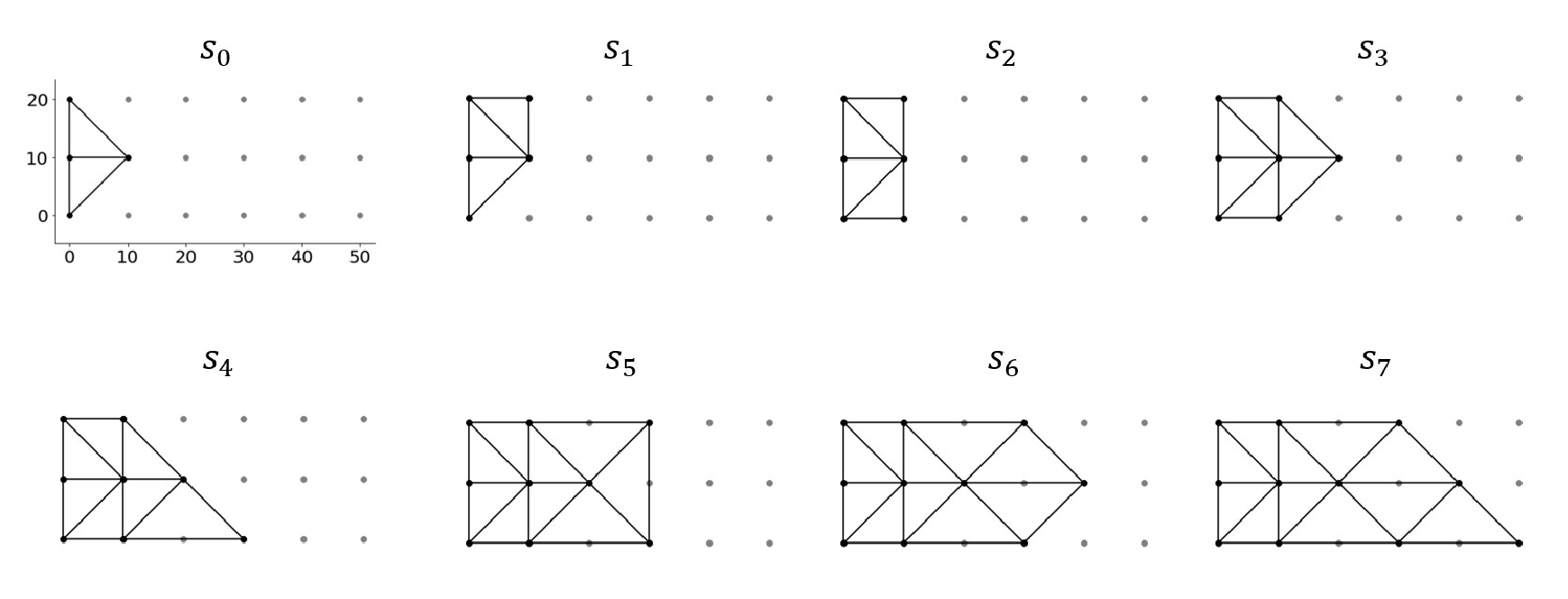}
\caption{Progressive construction - Cantilever case study: sequence of design configurations from the target optimal policy.\label{fig:cantilever_states}}
\end{figure*}

\begin{figure}[t]
\centering
\includegraphics[width=\linewidth]{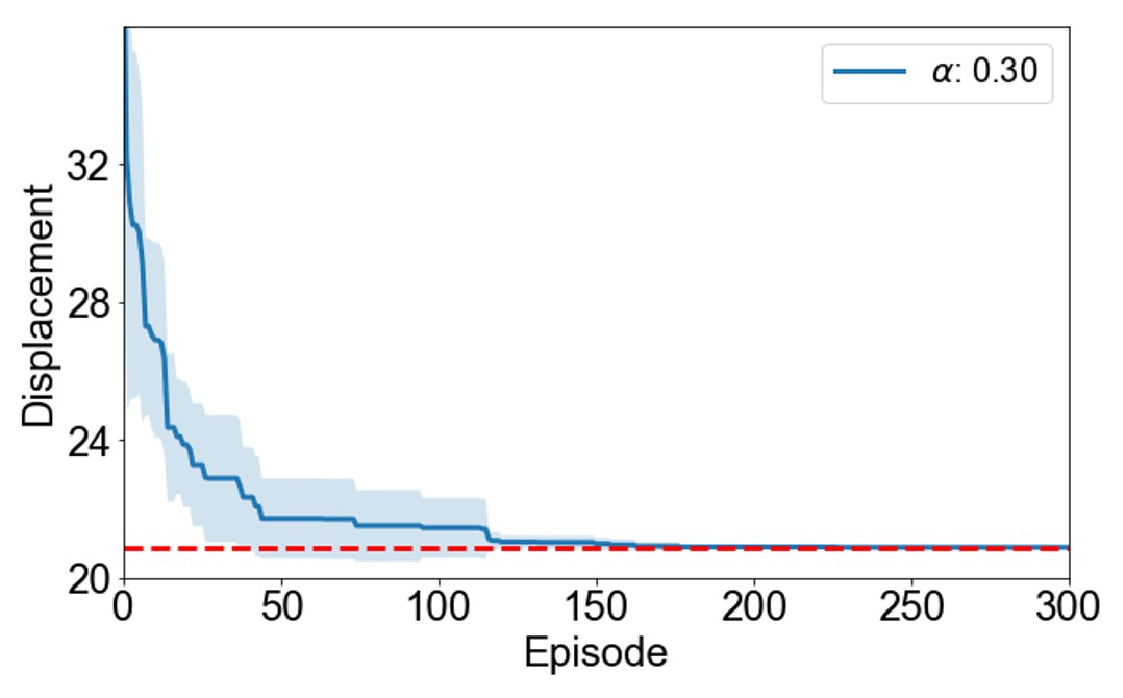}
\caption{Progressive construction - Cantilever case study: evolution of the design objective during training, shown as the average value (solid blue line) with its one-standard-deviation credibility interval (shaded blue area), and target global minimum (dashed red line). Results averaged over 10 training runs.\label{fig:cantilever_displacement_episode}}
\end{figure}

\begin{figure*}[t]
\centering
\includegraphics[width=\textwidth]{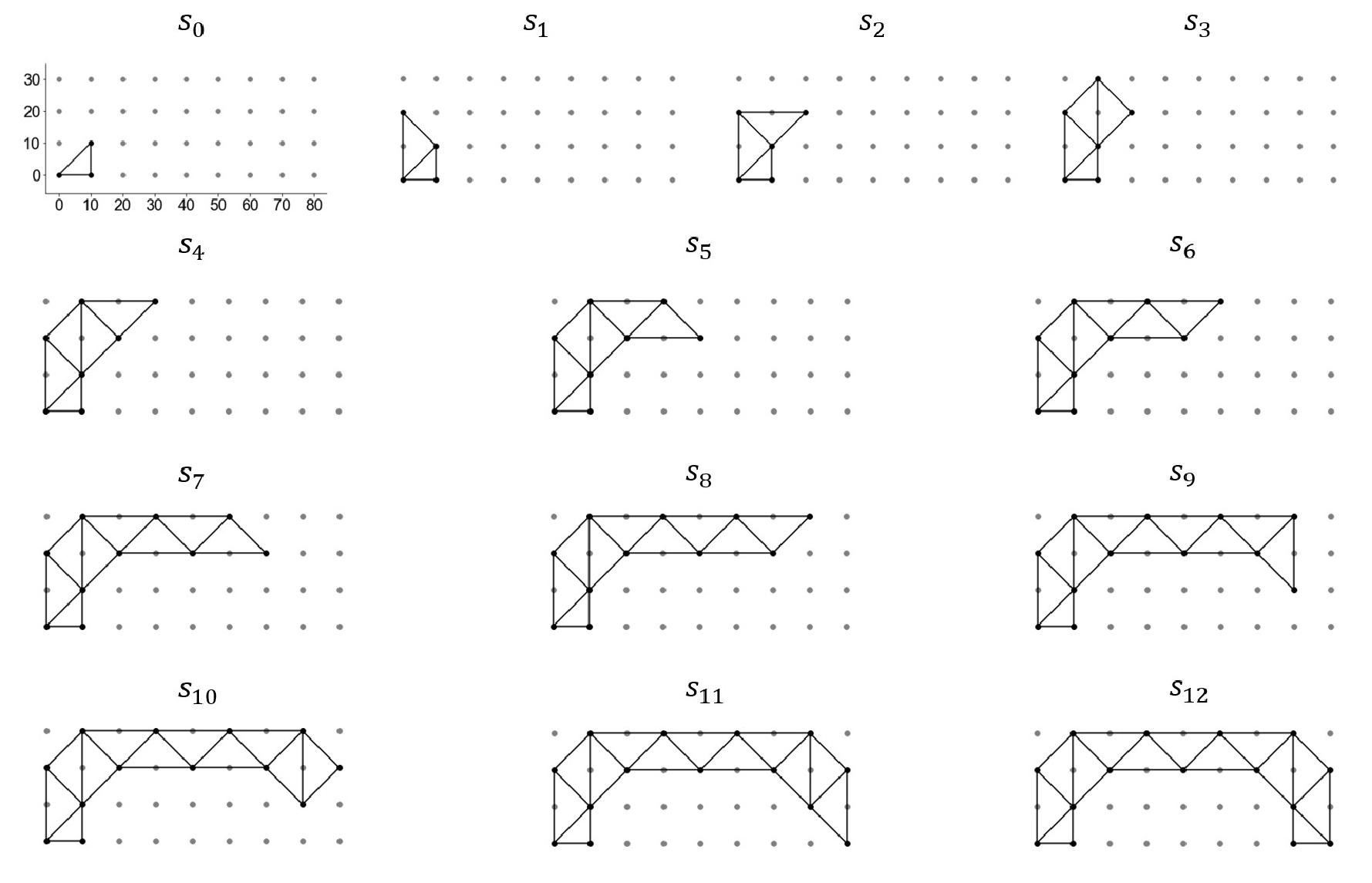}
\caption{Progressive construction - Bridge-like case study: sequence of design configurations from the target optimal policy.\label{fig:bridge_states}}
\end{figure*}

\begin{figure}[t]
\centering
\includegraphics[width=\linewidth]{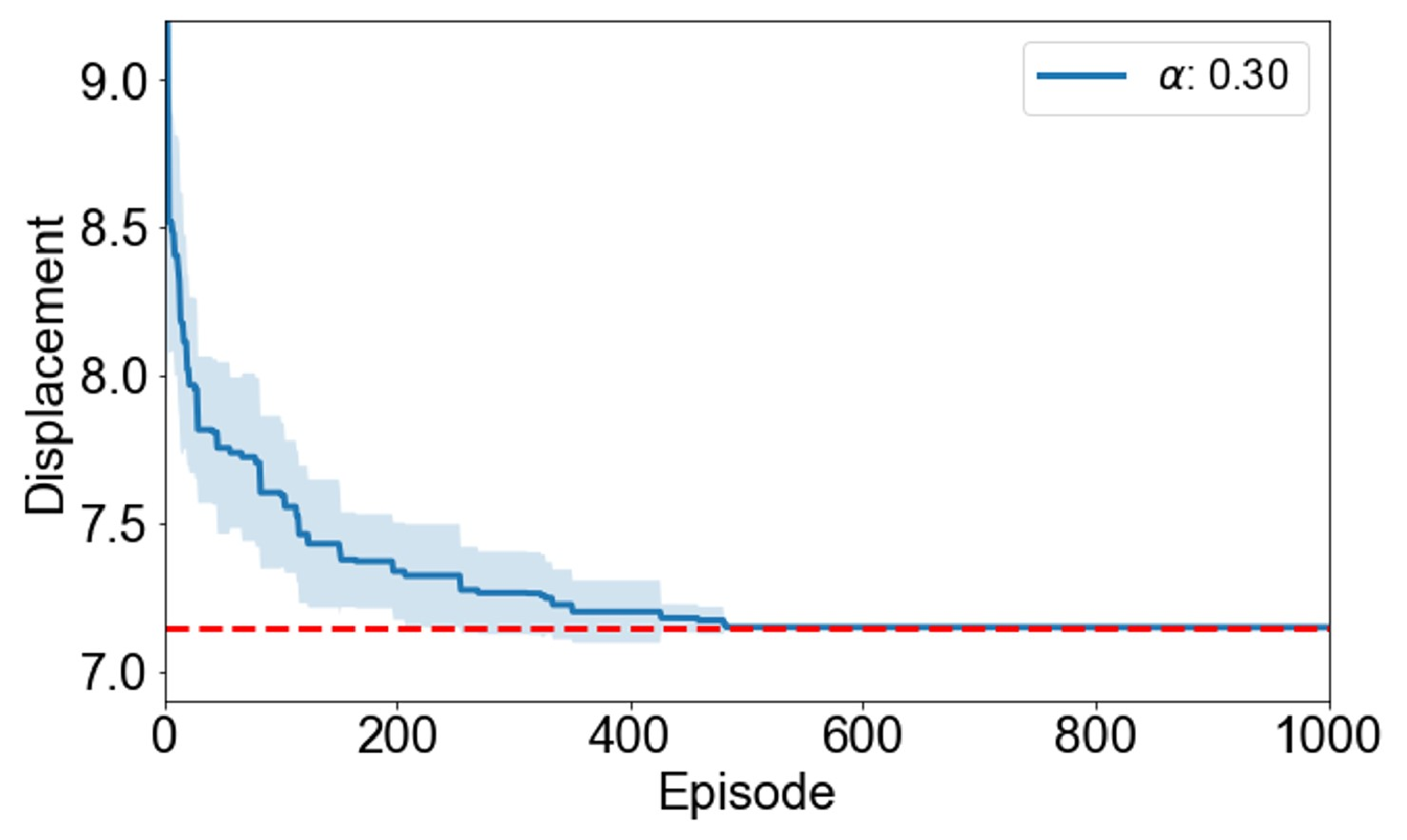}
\caption{Progressive construction - Bridge-like case study: evolution of the design objective during training, shown as the average value (solid blue line) with its one-standard-deviation credibility interval (shaded blue area), and target global minimum (dashed red line). Results averaged over 10 training runs.\label{fig:bridge_displacement_episode}}
\end{figure}


For the cantilever case study, we assign a domain size of $50\times20$, while for the bridge-like case study, we consider a larger domain of $80\times30$, which features a central passive area where FEs can not be connected. The sequence of optimal design configurations is shown in \fig\ref{fig:cantilever_states} for the cantilever beam, and in \fig\ref{fig:bridge_states} for the bridge-like structure. These optimal sequences have been synthesized from an exhaustive search, halted due to computational constraints after scanning $10,000,000$ and $1,200,000$ possible configurations, respectively. The maximum length of the individual elements has been constrained to comply with the typical fabrication, transportation, and on-site assembly limitations encountered in construction projects. This realistic constraint compels the algorithm to explore more detailed designs, avoiding trivial configurations that rely only a few long elements to reach the target node.

For the cantilever case study, the optimal configuration is synthesized $100\%$ of the time over 10 training runs. Using $\alpha = 0.3$, MCTS requires an average of 507 FE evaluations per training run, yielding an optimal displacement of $20.861$. It is worth noting how the algorithm identifies the optimal configuration by focusing on the most promising solutions, which feature more elements near the clamped side rather than near the free end (see $s_7$ in \fig\ref{fig:cantilever_states}). Refer to \fig\ref{fig:cantilever_displacement_episode} for the evolution of the attained design objective as the number of episodes increases during training.

Similarly, for the bridge-like case study, the MCTS policy synthesizes the optimal configuration $100\%$ of the time over 10 training runs. The evolution of states from the MCTS policy is identical to that of the target optimal policy, as shown in \fig\ref{fig:bridge_states}. Using $\alpha = 0.3$, the algorithm requires an average of 901 FE evaluations per training run, yielding an optimal displacement of $7.147$.  Finally, \fig\ref{fig:bridge_displacement_episode} presents the corresponding evolution of the attained design objective during training.

These case studies highlight the advantages of MCTS over Q-learning approaches for optimal design synthesis in large state spaces. In such cases, Q-learning struggles because it requires sufficient sampling of each state-action pair to build a Q-table that stores values for every possible pair, leading to exponential growth in memory and computational demands as the number of states increases. In contrast, MCTS dynamically builds a decision tree based on the most promising moves explored through simulation, focusing computational resources on more relevant parts of the search space. This selective exploration allows MCTS to handle large state spaces more efficiently than Q-learning, making it better suited to problems where direct enumeration of all state-action pairs is infeasible.

\subsection{Discussion}
\label{sez:discussion}

In the case studies used for comparison with \cite{Ororbia2023}, the proposed MCTS framework has been capable of synthesizing a near-optimal solution with significantly fewer FE evaluations. Case study 2 has shown the greatest reduction, requiring $76\%$ fewer evaluations. All examples have achieved a percentile score greater than $99.9\%$ when compared to the candidate solutions from the exhaustive search. In the first three case studies, the global optimal solution has been synthesized $100\%$ of the time. However, this was not the case for the last three. 

As seen in \tab\ref{tab:case_study_percentile_scores}, the lowest percentile score has been achieved in case study 4. A significant factor contributing to this result is the large number of decision nodes present in the first layer of the search tree. Although case study 4 features a smaller search space compared to case study 5, it exhibits a considerably larger branching factor at the initial layer, with 94 child nodes as opposed to 42 in case study 5. This increased number of initial options introduces a greater degree of complexity, slowing down the algorithm's convergence rate. Wider trees are more computationally intensive, as they potentially feature promising paths hidden among a multitude of branches. This aspect may lead to insufficient exploration of the branches departing from each node, resulting in less accurate value estimation. A straightforward solution to improve the approximation accuracy is to increase the number of episodes, thereby allowing each child node more visits. Another factor contributing to the lower percentile score in case study 4 is that the optimal solution is located in a section of the tree crowded with numerous lower-quality configurations. The UCT formula \eqref{eq:UCT} favors the exploitation of areas of the tree that, on average, yield good results. Consequently, the procedure tends to overlook tree branches that could potentially lead to the global optimum in favor of branches that consistently yield good configurations. The rationale is that areas of the tree yielding, on average, better solutions typically also contain the optimal configuration. However, this is not necessarily true for every problem, and it is not the case for case study 4. Even though reaching the global optimum is not precluded, this limitation of the UCT formula may slow convergence to the optimum. To address these drawbacks, the UCT equation could be modified to account for the standard deviation and the maximum value of the reward obtained from traversing the branch of child nodes, similarly to \cite{Schadd2008, Jacobsen2014}, as follows:
\begin{equation}
\label{eq:UCT_2}
\text{UCT}^{\text{new}}_j = (1-\alpha)\left[\frac{(1-\beta)v^\Sigma_j}{n_j}+\beta v^\text{best}_j\right] + \alpha \sqrt{\frac{2\log \sum_l n_l}{n_j}+\text{var}(v_j)},
\end{equation}
where, $v^\text{best}_j$ and $\text{var}(v_j)$ represent the maximum value and the variance of the reward gained by passing through the $j$-th child node, respectively, while $\beta$ is a hyperparameter that balances exploitation between tree areas consistently yielding good results and areas containing the best-seen state. This modified selection policy is less likely to be unfairly biased towards nodes with fewer children and promotes increased exploration.

The heuristic $\alpha$ parameter balances the exploitation and exploration terms in the UCT formula. From \fig\ref{fig:fig8b}, we observe that for both $\alpha = 0.1$ and $\alpha = 0.2$, MCTS converges early on local minima. This occurs because the first term in the UCT formula dominates the second term, preventing sufficient exploration of child nodes in the tree. While increasing the value of $\alpha$ mitigates this issue, the number of FE evaluations is observed to rise linearly with $\alpha$, as shown in \fig\ref{fig:example4_FEM_alpha}. However, the greater computational burden required by $\alpha=0.4$ and $\alpha=0.5$ results only in a limited improvement in the percentile score compared to $\alpha = 0.3$, as shown in \fig\ref{fig:example4_percentile}. Thus, $\alpha = 0.3$ has been chosen to balance a high percentile score with a low number of FE evaluations.

The cantilever and bridge-like case studies discussed in \sez\ref{sez:cantilever}, within the context of progressive construction, demonstrate the potential of the proposed MCTS framework to synthesize optimal designs in problems with very large state spaces. The (partial) exhaustive search has analyzed more than $10,000,000$ possible design configurations for the cantilever beam and $1,200,000$ for the bridge-like structure. Despite this, the optimal solution has been synthesized $100\%$ of the time, requiring only 507 FE evaluations per training run for the cantilever beam and 901 for bridge-like structure. In contrast, case study 4, which features a much smaller state space, has required significantly more FE evaluations without achieving the global optimum. This discrepancy is partly due to the fact that, although there are more layers in the trees of the progressive construction case studies, each layer is much narrower, allowing the algorithm to more easily identify promising branches. For instance, in the cantilever case, the first layer of the decision tree has 15 times fewer nodes compared to case study 4, resulting in much lower complexity.

One strength of MCTS over Q-learning \cite{Ororbia2021} and deep Q-learning \cite{Ororbia2023} approaches to optimal design synthesis is its ability to backpropagate reward signals to ancestor nodes in the tree more effectively. The difficulty deep Q-learning faces in performing the backpropagation step has also been mentioned in \cite{Ororbia2023}. Another limitation of Q-learning is that the reward signal is not continuously positive. Q-learning updates Q-values based on the difference between future and current reward estimates, adjusting only the values for the state-action pairs experienced in each step. These incremental updates can not take place before backpropagating information from the final stage to the intermediate design configurations. These drawbacks are not present in the MCTS framework because the reward is computed at the end of every episode, as is typical in Monte Carlo approaches, in contrast to temporal differences methods like Q-learning. Another key advantage of MCTS is that it builds a tree incrementally and selectively, exploring parts of the state space that are more promising based on previous episodes. This selective expansion is particularly advantageous in environments with extremely large or infinite state spaces, where attempting to maintain a value for every state-action pair (as in Q-learning) becomes infeasible. While we may not synthesize the absolute global optimum solution every time, we are able to achieve very high percentile scores. Most importantly, this framework can scale to large state spaces without significantly compromising the relative quality of the solutions.

\section{Conclusion}
\label{sez:conclusion}

This study has presented a comprehensive analysis of combining Monte Carlo tree search (MCTS) and generative grammar rules to optimize the design of planar truss lattices. The proposed framework has been tested across various case studies, demonstrating its capability to efficiently synthesize near-optimal (if not optimal) configurations even in large state spaces, with minimal computational burden. Specifically, we have compared MCTS with a recently proposed approach based on deep Q-learning, achieving significant reductions in the number of required finite element evaluations, ranging from $31\%$ to $76\%$ across different case studies. Moreover, two novel case studies have been used to highlight the adaptability of MCTS to dynamic and large state spaces typical of progressive construction scenarios. A critical analysis has been carried out to explain why the method has been able to get close to the global optimum without reaching it in some cases. Specifically, we have noted that this difficulty is not solely connected to the size of the state space but is due to the width of the tree and the adopted UCT formula. This formula encourages exploration of tree sections with the best average reward, potentially neglecting sections with lower average rewards that may contain the global optimum.

Compared to Q-learning, the proposed MCTS-based strategy has demonstrated two key advantages: ($i$) an improved capability of backpropagating reward signals, and ($ii$) the ability to selectively expand the decision tree towards more promising paths, thereby addressing large state spaces more efficiently and effectively.

The obtained results underscore the potential of MCTS not only in achieving high-percentile solutions but also in its scalability to larger state spaces without compromising solution quality. As such, this framework is poised to be a robust tool in the field of structural optimization and beyond, where complex decision-making and extensive state explorations are required. In the future, modifications to the UCT formula will be explored to address the occasional challenges in reaching the global optimum. Moreover, we foresee the possibility of exploiting this approach in progressive construction, extending beyond the domain of planar truss lattices.

\section*{Acknowledgements}

\noindent The authors of this paper would like to thank Ing. Syed Yusuf and Professor Matteo Bruggi (Politecnico di Milano) for the invaluable insights and contributions during our discussions.

\section*{Funding Data}
\noindent This work is partly supported by ERC advanced grant IMMENSE -- 101140720. (Funded by the European Union. Views and opinions expressed are however those of the authors only and do not necessarily reflect those of the European Union or the European Research Council Executive Agency. Neither the European Union nor the granting authority can be held responsible for them)\\[4pt]
\noindent Matteo Torzoni acknowledges the financial support from Politecnico di Milano through the interdisciplinary PhD grant ``Physics-Informed Deep Learning for Structural Health Monitoring''.

\appendix

\section{Parametric analysis of the $\alpha$ value}
\label{sec:parametric}

\begin{figure*}[t]
\centering
\includegraphics[width=1\textwidth]{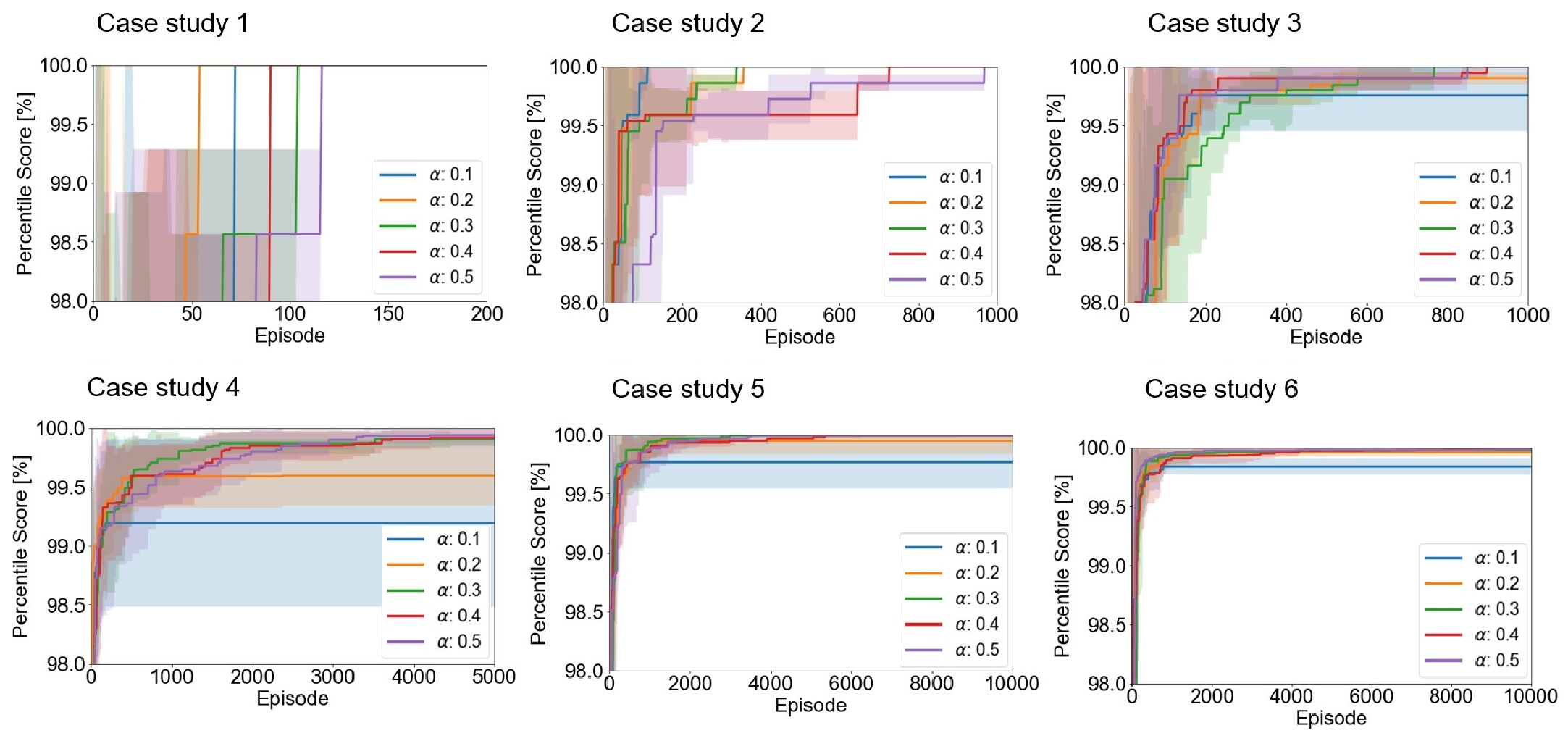}
\caption{Truss optimization - Case studies 1-6: impact of varying the $\alpha$ parameter on the attained percentile score relative to the exhaustive search space. For each value of $\alpha$, results are reported in terms of the evolution of the percentile score during training, shown as the average value with its one-standard-deviation credibility interval. Results averaged over 10 training runs.\label{fig:example_1_6_percentile_episode_alpha}}
\end{figure*}

In this Appendix, we provide results from a parametric analysis of the exploitation-exploration parameter $\alpha$ for case studies 1-6. Our modified UCT formula \eqref{eq:UCT} employs the $\alpha$ parameter to control the relative scaling of the exploitation and exploration terms, in contrast to the standard UCT formula, shown below:
\begin{equation}
\label{eq:UCT_3}
\text{UCT}_j = \frac{v^\Sigma_j}{n_j} + c \sqrt{\frac{2\log \sum_l n_l}{n_j}}.
\end{equation}
For $\alpha = 0.5$, the modified UCT formula in \eq\eqref{eq:UCT} coincides with \eq\eqref{eq:UCT_3} when $c = 1$. The choice of $c = 1$ is based on theoretical bounds that optimize the exploitation-exploration balance for the general case of the multi-armed bandit problem, where rewards are normalized between 0 and 1 \cite{Auer2002}. While this serves as a good starting point, determining the appropriate $\alpha$ value is challenging to ascertain a-priori, as it is case-dependent, as demonstrated in \fig\ref{fig:example_1_6_percentile_episode_alpha}. A small choice of $\alpha$ can heavily prioritise exploitation, as shown in \fig\ref{fig:example_1_6_percentile_episode_alpha} for case studies 3-6 with $\alpha = 0.1-0.2$, which plateau well below the global optimum. This is also summarized in \tab\ref{tab:case_studies_varying_alpha}, where we provide the percentile scores and the number of required FE evaluations at varying $\alpha$ for the six case studies. More FE runs means more unique terminal states and more branches explored in the tree.

\begin{table*}[ht]
\caption{Truss Optimization - Results for case studies 1-6: impact of varying the $\alpha$ parameter on the attained percentile score relative to the exhaustive search space, and on the number of finite element evaluations required to achieve a near-optimal or optimal policy. Results averaged over 10 training runs.\label{tab:case_studies_varying_alpha}}
\centering
\begin{tabular}{@{}llccccc@{}}
\toprule
 &  & $\alpha = 0.1$ & $\alpha = 0.2$ & $\alpha = 0.3$ & $\alpha = 0.4$ & $\alpha = 0.5$ \\ 
\midrule
\multirow{2}{*}{\textbf{Case Study 1}} 
 & \textbf{Percentile} & 100\% & 100\% & 100\% & 100\% & 100\% \\ 
 & \textbf{FE runs}        & 63     & 128    & 106    & 279    & 279    \\ 
\midrule
\multirow{2}{*}{\textbf{Case Study 2}} 
 & \textbf{Percentile} & 100\% & 100\% & 100\% & 100\% & 100\% \\ 
 & \textbf{FE runs}        & 173    & 545    & 517    & 978    & 988    \\ 
\midrule
\multirow{2}{*}{\textbf{Case Study 3}} 
 & \textbf{Percentile} & 99.75\% & 99.90\% & 100\% & 100\% & 100\% \\ 
 & \textbf{FE runs}        & 268     & 623     & 961    & 966    & 966    \\ 
\midrule
\multirow{2}{*}{\textbf{Case Study 4}} 
 & \textbf{Percentile} & 99.19\% & 99.59\% & 99.90\% & 99.91\% & 99.94\% \\ 
 & \textbf{FE runs}        & 217     & 710     & 1672   & 2800   & 3433   \\ 
\midrule
\multirow{2}{*}{\textbf{Case Study 5}} 
 & \textbf{Percentile} & 99.77\% & 99.95\% & 99.99\% & 99.99\% & 99.99\% \\ 
 & \textbf{FE runs}        & 586     & 2699    & 7235   & 9717   & 9739   \\ 
\midrule
\multirow{2}{*}{\textbf{Case Study 6}} 
 & \textbf{Percentile} & 99.83\% & 99.96\% & 99.98\% & 99.98\% & 99.99\% \\ 
 & \textbf{FE runs}        & 770     & 3270    & 7931   & 9204   & 9357   \\ 
\bottomrule
\end{tabular}
\end{table*}

\begin{table*}[t]
\centering
\caption{Bridge-like case study: timing breakdown of the MCTS phases during a pilot training session of 1000 episodes.\label{tab:mcts_phase_breakdown}}
\begin{tabular}{l c c}
\toprule
\textbf{MCTS Phase}      & \textbf{Time (seconds)} & \textbf{Percentage} \\ 
\midrule
Selection                & 0.91                 & 0.09\%                      
\\ 
Expansion                & 163.08                & 16.89\%                     
\\ 
Simulation               & 801.00               & 82.97\%                     
\\ 
Backpropagation          & 0.38                 & 0.039\%                      
\\ 
Other                    & 0.01                  & 0.01\%                      
\\ 
\midrule
\textbf{Total Elapsed}   & \textbf{965.3872}       & \textbf{100\%}               \\ 
\bottomrule
\end{tabular}
\end{table*}

\section{Computational cost analysis}
\label{sec:computation_cost}

In this appendix, we provide an overview of the computational burden associated with a pilot MCTS training of 1000 episodes for the bridge-like case study. The timing analysis is summarized in \tab\ref{tab:mcts_phase_breakdown}, reporting the computational time taken by each MCTS phase. The simulation phase dominates the computational time, accounting for 82.97\% of the total execution time, followed by the expansion phase at 16.89\%. In contrast, the selection and backpropagation phases require significantly less time, contributing 0.09\% and 0.04\%, respectively. The remaining operations, collectively termed as ``Other'', take up a minimal 0.01\% of the total execution time.

We have identified that the most computationally demanding task is not the FE analysis itself, but rather the frequent execution of relatively simple geometric checking functions that determine whether a new configuration violates geometric constraints. Specifically, a function that checks whether a line passes over an active node has been called 14,559,543 times during execution. This function, invoked primarily during the child node population process, has been responsible for a cumulative execution time of 438.77 seconds, representing 45.45\% of the total runtime. These calls occurred in both the expansion and simulation phases, significantly contributing to the overall computational load despite the function's simplicity. Overall, in the simulation phase, the algorithm must populate more layers of children nodes than in the expansion phase, therefore contributing to its higher computational cost.

\bibliographystyle{asmejour}

\end{document}